\documentclass[prl, showpacs, twocolumn, notitlepage, nofootinbib, superscriptaddress
]{revtex4-1}
\usepackage{amsfonts, amsmath, amssymb}
\usepackage[english]{babel}
\usepackage{comment}
\usepackage[dvips]{graphicx}
\usepackage{latexsym}

\usepackage[normalem]{ulem}
\usepackage{ifthen}
\newboolean{ARXIV}
\setboolean{ARXIV}{true}
\newcommand{\arxiv}[1]{\ifthenelse{\boolean{ARXIV}}{ [#1]}{}}

\usepackage[T1]{fontenc}
\usepackage{pslatex}
\usepackage{color}

\newcommand{\be}{\begin{equation}}
\newcommand{\ee}{\end{equation}}
\newcommand{\ba}{\begin{aligned}}
\newcommand{\ea}{\end{aligned}}

\begin{document}

\title{Armchair graphene nanoribbons: $\mathcal{PT}$-symmetry breaking \\ and exceptional points without dissipation}

\author{Maurizio Fagotti}
\author{Claudio Bonati}
\affiliation{Dipartimento di Fisica, Universit\`a di Pisa and INFN
, Largo Pontecorvo 3, I-56127 Pisa, Italy.}

\author{Demetrio Logoteta}
\author{Paolo Marconcini}
\author{Massimo Macucci}
\affiliation{Dipartimento di Ingegneria dell'Informazione, Universit\`a di Pisa, Via Caruso 16, I-56122 Pisa, Italy.}

\begin{abstract}
We consider a single-layer graphene nanoribbon with armchair edges and with a longitudinally constant external 
potential, pointing out that it can be described by means of an effective non-Hermitian Hamiltonian. 
We show that this system has some features typical of dissipative systems, namely the presence of exceptional points and of $\mathcal{PT}$-symmetry breaking, although it is not dissipative. 
\end{abstract}


\maketitle

Since its isolation in 2004 \cite{Novoselov04}, graphene has attracted a significant interest in the condensed matter 
community because of its unique electronic properties, the most astonishing one being the presence in its low-energy 
spectrum of two massless Dirac modes. These massless fermionic excitations allowed the experimental observation 
of exotic phenomena, such as the Klein paradox and the Zitterbewegung (see \emph{e.g.} \cite{CastroNeto} for a review),
theoretically studied in the context of quantum electrodynamics long before. 

In this letter we analyze some properties of single-layer graphene nanoribbons with armchair edges in an
external potential, pointing out a connection with the recently developed theory of $\mathcal{PT}$-symmetric 
non-Hermitian Hamiltonians in quantum mechanics. 

The systematic study of $\mathcal{PT}$-symmetric Hamiltonians was initiated by the seminal paper \cite{Bender98}, 
in which analytical and numerical hints were presented to explain the reality of the spectra of some non-Hermitian 
Hamiltonians. If the $\mathcal{PT}$ symmetry is realized in the spectrum, \emph{i.e.} if every eigenstate of the 
Hamiltonian is also an eigenstate of the $\mathcal{PT}$ operator, it is simple to show that the energy spectrum has 
to be real. However, since $\mathcal{PT}$ is an anti-linear operator, the symmetry can be spontaneously broken  
(for some enlightening examples and a review of the main results obtained in this field see \emph{e.g.} \cite{Bender07}) 
and EPs (exceptional points) can appear. Exceptional points, \emph{i.e.} points for which two (or more) eigenvalues 
coalesce and the Hamiltonian is non-diagonalizable \cite{Kato}, are a typical feature of non-Hermitian Hamiltonians 
(see \emph{e.g.} the reviews \cite{EP}) with no Hermitian counterpart and they have been shown to produce experimentally 
observable effects \cite{mw,EP_exp1, EP_exp2}.

To our knowledge, all the previously proposed physical examples of systems governed by non-Hermitian 
$\mathcal{PT}$-symmetric Hamiltonians involve dissipative systems, the main emphasis being on microwave cavities \cite{mw},
optical lattices \cite{optic_th, optic_exp} and lasers \cite{laser}. We show that, because of the spinorial nature of the 
wave function, some properties of graphene nanoribbons can be described by means of an effective non-Hermitian 
$\mathcal{PT}$-symmetric Hamiltonian, although there is no dissipation. We give numerical evidence for the 
$\mathcal{PT}$-symmetry breaking and provide an order parameter. Finally we study the behavior of eigenmodes 
and eigenfunctions in the neighborhood of exceptional points.

\paragraph{The model and the notations.} We consider a nanoribbon section with open boundary conditions along the longitudinal $x$ direction, armchair edges, and $N$ dimer lines across its width. The transverse distance (along the $y$ direction) between 
the first line of lattice points not occupied by carbon atoms at the bottom edge and the analogous one at the upper 
edge of the ribbon (\emph{i.e.} the ``effective width'' of the ribbon) is equal to $L=(N+1)a/2$, with $a$ the graphene 
lattice constant. The wave function is equal to 
\be
\psi (\vec r)=
\sum_{\vec R_A}\psi_A (\vec R_A)\varphi(\vec r -\vec R_A)+
i \sum_{\vec R_B}\psi_B (\vec R_B)\varphi(\vec r -\vec R_B)\, ,
\ee
where the $\varphi(\vec r)$'s are the orthonormalized $2 p^z$ atomic orbitals of carbon, $\vec R_A$ and $\vec R_B$ 
are the positions of the atoms of the two sublattices $A$  and $B$ of graphene, and
\be
\vec{\Psi}(\mathbf{r})=
\begin{pmatrix}
\psi_A (\mathbf{r})\\
\psi_B (\mathbf{r})
\end{pmatrix}=
e^{-iKy}\vec{\chi}^{\vec K}(x,y)-e^{iKy}\vec{\chi}^{\vec K'}(x,y)\, ,
\ee
with $K=4\pi/(3a)$ and $\vec{\chi}^{\vec K}$, $\vec{\chi}^{\vec K'}$ the $2$-spinors corresponding to the two 
inequivalent Dirac points $\vec K$ and $\vec K'$. The external potential $V(x,y)$ is assumed to vary only 
in the transverse direction. Aside from its intrinsic interest, evaluation of the eigenfunctions and eigenvalues for such 
a potential represents the first step in conductance calculations for a generic potential (varying in both spatial 
directions) performed by means of scattering matrix methods. Most existing calculations of transport in graphene 
based on such methods deal instead with a constant transverse potential \cite{wurm,li}.

We can expand $\vec{\chi}^{\vec K}$, $\vec{\chi}^{\vec K'}$ in plane waves along the $x$ direction and write 
$\vec{\chi}^{\vec K}(x,y)=e^{i k_x x} \vec{\varphi}^{\vec K}(y)$ and
$\vec{\chi}^{\vec K'}(x,y)=e^{i k_x x} \vec{\varphi}^{\vec K'}(y)$.
The Dirac equation can be written in the form (see \emph{e.g.} \cite{bc,CastroNeto})
\be \label{Dirac0}
\left\{
\ba
& \big(f(y)+\sigma_x k_x-i\sigma_y\partial_y\big)\vec{\varphi}^{\vec K}(y)=0 \\
& \big(f(y)+\sigma_x k_x+i\sigma_y\partial_y\big)\vec{\varphi}^{\vec K'}(y)=0
\ea
\right.
\ee
where we introduced the shorthand $f(y)=[V(y)-E]/v_F$, with $v_F$ the Fermi velocity (we use $\hbar=1$ in order to 
simplify the notations), and the armchair boundary conditions read
\be \label{bc0}
\vec{\varphi}^{\vec K}(0)=\vec{\varphi}^{\vec K'}(0)\ ,\quad
\vec{\varphi}^{\vec K}(L)=e^{2ikL}\vec{\varphi}^{\vec K'}(L) .
\ee
The $k_x$ values for which this system admits non trivial solutions are the longitudinal momenta allowed in the nanoribbon.  
The problem of Eqs.~\eqref{Dirac0}-\eqref{bc0} can be rewritten in a more convenient form introducing, for $y\in[-L,L]$, 
the $2$-spinor 
\be
\vec\varphi(y)=\begin{cases}
\vec\varphi^{\vec K}(y+L)&y\in[-L,0]\\
e^{2i KL}\,\vec\varphi^{\vec K'}(L-y)&y\in]0,L]
\end{cases}
\ee
and defining $\bar{f}(y)=f(L-|y|)$ and $H^{k_x}(y)=\bar{f}(y)\sigma_y-ik_x\sigma_z$. We obtain 
\be \label{Dirac1}
\ba
i\partial_y\vec{\varphi}(y)=H^{k_x}(y)\vec{\varphi}(y),\quad \vec{\varphi}(L)=e^{2 i K L}\vec{\varphi}(-L),
\ea
\ee
which is formally equivalent to a Schr\"odinger equation for the non-Hermitian Hamiltonian $v_F H^{k_x}$, if we interpret 
$y/v_F$ as the time. 

\paragraph{Symmetries.}
From Eq.~\eqref{Dirac1} we deduce a simple result on the degeneration of the $k_x$ modes. If we denote by $U(y)$ the 
time evolution operator associated to a given eigenvalue $k_x$, \emph{i.e.} $\vec{\varphi}(y)=U(y)\vec{\varphi}(-L)$ 
with $\vec\varphi(y)$ in the corresponding eigenspace, the boundary condition can be written as 
$U(L)\vec{\varphi}(-L)=e^{2iKL}\vec{\varphi}(-L)$. By using the explicit form of $H^{k_x}$ it is easy to check that 
$U^{-1}(L)=\sigma_x U(L)\sigma_x$, from which it follows that 
$U(L)\sigma_x\vec{\varphi}(-L)=e^{-2iKL}\sigma_x\vec{\varphi}(-L)$. 
If $\exp(4iKL)\ne 1$, there cannot be  eigenvectors of $U(L)$ other than $\vec\varphi(-L)$ and $\sigma_x\vec\varphi(-L)$, 
and hence just one independent eigenmode corresponds to each $k_x$. From now on we consider lengths $L$ for which 
this condition is verified, \emph{i.e.} nanoribbons that are semiconducting in the absence of an external potential also 
when edge relaxation is neglected. We denote by $\vec{\varphi}_{k_x}(y)$ the eigenmode associated to $k_x$. 
From the relations
\be\label{z4}
\ba
&\vec\varphi_{k_x^*}(y)\sim\sigma_x\bigl(\vec\varphi_{k_x}(-y)\bigr)^*&&
&\vec\varphi_{-k_x^*}(y)\sim\sigma_z\bigl(\vec\varphi_{k_x}(-y)\bigr)^*
\ea
\ee
it follows that if $k_x$ is in the spectrum then there are also $k_x^*$, $-k_x^*$ and $-k_x$. Thus the spectrum has
a $Z_2\times Z_2$ symmetry.

\begin{figure}[t]
\vspace{1mm}
\scalebox{0.3}{\includegraphics*{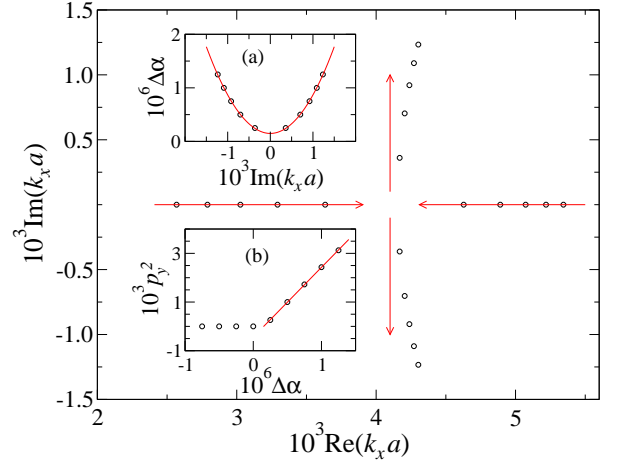}}
\caption{Coalescing of two eigenvalues along the real line for $\epsilon=1/5$; the arrows denote the direction of 
increasing $\alpha$ values ($\Delta\alpha\equiv\alpha-0.060953$). In the inset (a) the imaginary parts of the coalescing 
eigenvalues are shown together with a fit of the form $\alpha=c_1 + c_2 (\mathrm{Im}\, (k_x a))^2$. 
The inset (b) shows the behavior of the square of the transverse momentum together with a linear fit.}
\label{fig_1}
\end{figure}

To reveal the $\mathcal{PT}$ symmetry of this problem it is convenient to take the square of Eq.~\eqref{Dirac1} and 
project $\vec{\varphi}$ on the eigenstates of $\sigma_y$. If we denote these projections by
$\phi_{\pm}$, they satisfy the equations 
\be\label{Dirac_square}
\Big(\hat{p}_y^2-\bar{f}^{\,2}\mp i (\partial_y \bar{f})\Big)\phi_\pm(y)=-k_x^2\phi_{\pm}(y),
\ee
which are clearly invariant under the $\mathcal{PT}$ transformation, being the action of the operators 
$\mathcal{P}$ and $\mathcal T$ defined by $p_y\rightarrow-p_y$, $y\rightarrow -y$ and $p_y\rightarrow-p_y$, 
$y\rightarrow y$, $i\rightarrow -i$, respectively.
If the $\mathcal{PT}$ symmetry is unbroken then $k_x$ has to be real or imaginary; complex conjugate (intended 
here as a number with nonzero real and imaginary part) pairs appear in the spectrum only if this symmetry is broken. 

We explicitly notice that if the Schr\"odinger equation is used instead of the Dirac one, the equation corresponding 
to Eqs.~\eqref{Dirac0} is similar to Eq.~\eqref{Dirac_square} but with an Hermitian left hand side, so that all the 
$k_x$ values have to be real or imaginary.

In the presence of spontaneously broken symmetries it is customary to look for an order parameter, \emph{i.e.} an 
observable that vanishes when the symmetry is realized in the spectrum, a non zero value signaling the symmetry breaking. 
We point out that the mean value of the transverse momentum  
\be\label{py_def}
p_y=\left.\int_0^L \vec{\Psi}(y)^{\dag}(-i\partial_y)\vec{\Psi}(y)\mathrm{d}y\right/
\int_0^L \vec{\Psi}(y)^{\dag}\vec{\Psi}(y)\mathrm{d}y  
\ee
satisfies this requirement. This can be easily proved exploiting the symmetries of our problem: from Eqs.~\eqref{z4} and 
because $|\vec{\varphi}(y)|^2$ is $\mathcal{P}$-even and $\vec\varphi(-y)^T\sigma_z \vec\varphi(y)$ is constant, it 
follows that the numerator of Eq.~\eqref{py_def} vanishes if $k_x^2$ is real. 
We checked numerically that the transverse momentum $p_y$ is different from zero when $k_x$ is complex 
($p_y$ appears to vanish also for complex $k_x$ values only when the potential satisfies $V(y)=V(L-y)$).
Thus, in this system the realization of the $\mathcal{PT}$ symmetry in the spectrum is related to the 
value of the transverse momentum, which is an order parameter for the $\mathcal{PT}$ symmetry breaking. This 
is completely analogous to what happens in QCD with compact dimensions, when charge conjugation can get
spontaneously broken and the related order parameter is the baryon current in the compactified direction \cite{lpp}.

Finally, it is worth mentioning that the transformation that reverses the sign of $\bar f(y)$ (\emph{i.e.} $\sigma_z$) 
is unitary and independent of $p_y$, and hence  $|\vec\varphi(y)|^2 $ is invariant for simultaneous flipping of the signs 
of the potential and of the energy. This is just a manifestation of the chiral symmetry of the system.  As long as  
$\bar f(y)$ does not have a definite sign, heuristic arguments based on that symmetry indicate the presence of 
eigenfunctions localized around the  minima as well as eigenfunctions localized around the maxima of the potential; 
the former (latter) ones are expected to describe particles with positive (negative, \emph{i.e.} of opposite sign 
with respect to $k_x$) group velocities. These observations provide a simple argument for the existence of some 
singular behavior: by varying the energy, eigenmodes with opposite group velocities coalesce in a non-analytic way, 
because they cannot combine in a two-dimensional eigenspace. 

\paragraph{$\mathcal{PT}$-symmetry breaking and exceptional points} 
If $f(y)$ is constant, Eq.~\eqref{Dirac1} can be analytically solved: all the $k_x$ values are real or imaginary 
and all the eigenstates of the Hamiltonian (their projections on the eigenstates of $\sigma_y$, to be precise)
are also eigenstates of the $\mathcal{PT}$ operator. This is no longer true when an external potential
with non-trivial $y$ dependence is present and we now provide numerical evidence that in this system 
$\mathcal{PT}$-symmetry can get spontaneously broken. As a simple example we use the Lorentzian shaped potential  
\be \label{test_case}
f(y)\,a=\epsilon -\frac{125}{250 + \alpha^2 ((y/a) - 150)^2},  
\ee
with $\alpha, \epsilon \in\mathbb{R}$. 
The general qualitative features are however independent of the particular potential adopted.
We choose $L=500a$ as the effective width of the nanoribbon.

It is simple to (numerically) check that by varying the parameters ($\alpha$ and $\varepsilon$ in our example) 
two different phenomena can occur:
\begin{itemize}
\item[A)] the number of the real 
eigenvalues varies but the number of the complex ones is preserved;
\item[B)] the number of the complex eigenvalues changes.
\end{itemize}
In case A) a couple of real eigenvalues turns into a couple of imaginary ones (or vice versa); the 
$Z_2\times Z_2$ symmetry then implies the existence of some values of the parameters for which $k_x=0$ is a 
doubly degenerate eigenvalue. However, we are assuming $\exp (4iKL)\ne 1$, and hence only one independent eigenfunction 
is associated to each eigenvalue; as a consequence, this point of the parameter space is an exceptional point. 
Case B) is completely analogous: two real or imaginary eigenvalues coalesce and become complex (see 
Fig.~\ref{fig_1}). However, while in case A) the $\mathcal{PT}$ symmetry is unbroken (or broken, if the number 
of complex eigenvalues is different from $0$) irrespective of the EP, in case B) the EP is associated to 
$\mathcal{PT}$ symmetry breaking. Clearly this kind of EP appears in a specular way both in the upper (right) 
and lower (left) half-plane, because of the $Z_2\times Z_2$ symmetry.

\paragraph{Behavior near EPs.} 
If we are not at an exceptional point, the variations of the eigenvalues and eigenvectors 
are linear in the variation $\delta\bar{f}$ of $\bar f$; in particular it can be shown that
\be
\int_{-L}^{L}\mathrm{d}y \vec\varphi^T_{k_x}(-y)\frac{\delta \bar{f}}{\delta k_x}\sigma_x\vec\varphi_{k_x}(y)=
-\int_{-L}^{L}\mathrm{d}y \vec\varphi^T_{k_x}(-y) \vec\varphi_{k_x}(y)\, .
\ee  
The group velocity $v^g_x$ in the $x$ direction is obtained in the special case of an energy shift and is given by
\be
v^g_x\equiv \frac{\delta E}{\delta k_x}=v_F \frac{\int_{-L}^{L}\mathrm{d} y\, \vec\varphi^T_{k_x}(-y)
\vec\varphi_{k_x}(y)}{\int_{-L}^{L}\mathrm{d}y\, \vec\varphi^T_{k_x}(-y) \sigma_x\vec\varphi_{k_x}(y)}\, .
\ee
The velocity defined as above has a clear physical interpretation for real modes, for which it is real; however 
also complex group velocities can provide interesting information in many physical systems (see \emph{e.g.} 
\cite{Complex_v}). 

\begin{figure}[t]
\vspace{1mm}
\scalebox{0.3}{\includegraphics{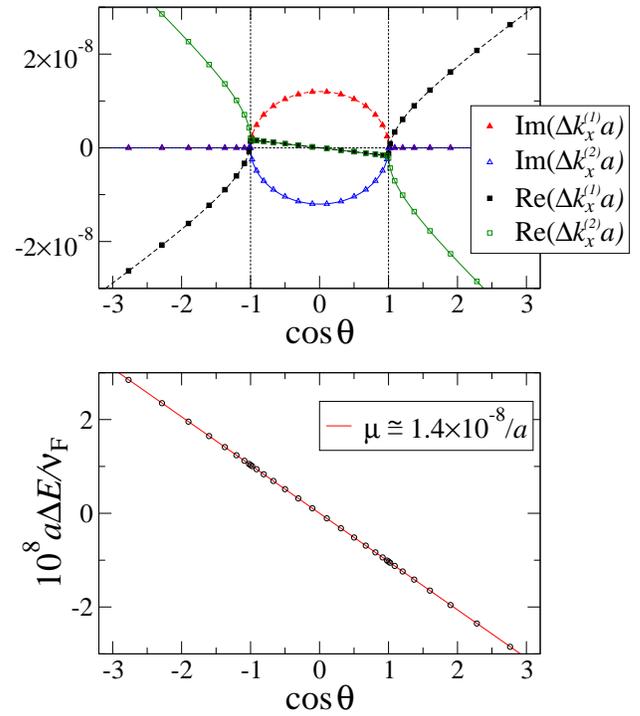}}
\caption{The momenta (\emph{top}) $\Delta k_{x}^{(1,2)}\approx k_{x}^{(1,2)}-0.199894795/a$ and the energy (\emph{bottom}) 
$\Delta E\approx E+0.200651930\,v_F/a$ of two coalescing modes as functions of $\cos\theta$, estimated from 
$(\det R-2)/\sqrt{1-\det R}$ (see Eqs.~\eqref{eq:mixing0} and \eqref{eq:mixing}), for the potential \eqref{test_case} with 
$\alpha=\pi/2$ and $\epsilon = -a\,E/v_F$. The lines in the upper graph follow the prediction 
\eqref{eq:modesenergy}: the asymptotic velocities $v_\pm$ are obtained by considering 
$\Delta E\gtrsim 2\cdot 10^{-5}\,v_F/a$ and  $\mu$  obtained from the linear fit of the energy versus $\cos\theta$ shown in 
the bottom graph.}
\label{fig_modes}
\end{figure}

At an EP, $\int_{-L}^{L}\mathrm{d}y\,\vec\varphi^T_{k_x}(-y) \vec\varphi_{k_x}(y)$ vanishes for some $k_x$ and in the 
neighborhood of an EP corresponding to the eigenvalue $k_x^{EP}$ we find
\be  \label{eq:criticalbehavior} 
\ba
\delta E\cong \frac{(k_x-k_x^{EP})^2}{2\mu}, \quad 
\mu=\frac{1}{v_F}\frac{\int_{-L}^{L}\mathrm{d} y\, \vec\varphi^T_{k_x}(-y)\sigma_x \vec\varphi_{k_x}(y)}{
\frac{\delta}{\delta k_x}\int_{-L}^{L}\mathrm{d} y\, \vec{\varphi}^T_{k_x}(-y) \vec\varphi_{k_x}(y)}\,. 
\ea
\ee
This is nothing but the well known square root behavior, in the neighborhood of an EP \cite{Kato, EP}, of two 
coalescing eigenvalues as a function of the external parameters (see the inset (a) of Fig.~\ref{fig_1}). 
 Near an EP we can then factorize the Hilbert space as the product of the $2d$ space of the collapsing 
eigenfunctions, for which the $k_x$'s rapidly change with a small shift in $E$ (fast modes), and the span of the other 
modes (slow modes), which can be assumed as fixed in the neighbourhood of the EP. Here we are assuming that eigenfunctions 
associated to different exceptional points do not mix; in the considered numerical examples we checked that this 
assumption is indeed true. Thus:
\be\label{eq:mixing0}
\begin{pmatrix}
\vec\varphi_{k_x^{ (1)}}\\\vec\varphi_{k_x^{(2)}}
\end{pmatrix}\approx R\begin{pmatrix}
\vec\varphi_{+}\\\vec\varphi_{-}
\end{pmatrix}\, ,
\ee  
where $k_x^{(1)}, k_x^{(2)}$ are the two coalescing eigenvalues, $R$ is a $2\times 2$ matrix and $\vec{\varphi}_{\pm}$ 
are the two initial states. Notice that if the difference between one or more eigenvalues and the coalescing ones is 
$\delta k_x\lesssim \mu$, they can mix together. It turns out that if they are quasi-degenerate with the coalescing 
eigenmodes ($\delta k_x\ll \mu$) the mixing between them is just a rotation, that is irrelevant for the features that 
we are going to describe. As long as we are not at the EP, we can choose the normalization
$\int_{-L}^L\mathrm d \varphi \vec\varphi_{k_x^\prime}^T(-y)\vec\varphi_{k_x}(y)=\delta_{k_x k_x^\prime}$, so 
that $R R^T=\mathrm I$. For the sake of simplicity, in the following we restrict to the case of 
$\delta\bar{f}=-\delta E/v_F$ and EPs on the real axis. The qualitative results obtained are nevertheless of 
general validity.
Before the EP is reached, the group velocities of the coalescing real eigenmodes are opposite in sign, as shown 
by Eq.~\eqref{eq:criticalbehavior} and indicated by the subscript $\pm$ in Eq.~\eqref{eq:mixing0}, and from 
Eqs.~\eqref{z4} it follows that $R^\ast = \sigma_z R \sigma_z$. After the EP is crossed, the eigenvalues as 
well as $\delta E/\delta k_x$ become complex conjugate and from Eqs.~\eqref{z4} it follows that $R^\ast = \pm 
\sigma_x R \sigma_z$. Relaxing the normalization condition, we can parametrize $R$ as follows:
\be\label{eq:mixing}
R\sim \mathrm I+e^{i\theta} \sigma_y\, .
\ee  
The domain of definition of  the parameter $\theta$ is $]\infty i ,0 i]\cup [0,\pi]\cup
[\pi+0i,\pi+\infty i[$. The $k_x$ values are real if $\mathrm{Im}\,\theta$ is different from $0$ and complex if $\theta$ 
is real; the larger $\mathrm{Im}(\theta)$ the further apart the modes $k_x^{(1,2)}$ are.  
When $\theta=0$ the eigenfunctions $\vec\varphi_{k_x^{(1,2)}}$ are linearly dependent, so this value corresponds to an 
exceptional point. We rewrite Eq.~\eqref{eq:criticalbehavior} in terms of the parameter $\theta$ in the simplest case 
in which the  term $\Delta\equiv\int_{-L}^L\mathrm d y \vec\varphi_{+}(-y)^T\sigma_x\vec\varphi_{-}(y)$, which appears 
in \(\mu\), is negligible 
\be\label{eq:modesenergy}
\left\{
\ba
&k_x^{(1,2)} \approx k_x^0+2\mu \frac{(v_+^{-1}+v_-^{-1})\cos\theta
\mp i(v_+^{-1}-v_-^{-1})\sin\theta}{(v_+^{-1}-v_-^{-1})^2}\\
&
E\approx E_0-\frac{4 \mu }{(v_+^{-1}-v_-^{-1})^2}\cos\theta\, ,
\ea
\right.
\ee
where $k_x^0$ and $E_0$ are constants and $v_\pm$ are the group velocities associated to the eigenfunctions 
$\vec\varphi_\pm$. The condition $\Delta\approx 0$ is found for example when one eigenfunction is localized around 
the minima and the other around the maxima of the potential. The approximate Eq.~\eqref{eq:modesenergy} is accurate 
only in a neighborhood of the EP with $\theta\approx 0$ but, if $\mu$ is much less than the energy scale in which 
the Hilbert space factorization remains valid, then the value $\theta=\pi$ of Eq.~\eqref{eq:mixing} corresponds 
actually to another EP, and the previous approximation is good in the whole interval $0\le \theta\le \pi$. This happens 
when two real eigenvalues collide, become complex and then come back on the real axis, the two EPs being sufficiently 
close to each other. In this case Eq.~\eqref{eq:modesenergy} captures the whole out-of-axes ``motion'' of the eigenvalues.
In Fig. \ref{fig_modes} the results predicted by Eqs.~\eqref{eq:mixing} and \eqref{eq:modesenergy} are checked against 
numerical data and the agreement appears to be excellent. After crossing both exceptional points, the eigenfunctions 
almost return to the starting ones; observe, moreover, that the energy scale of the phenomenon shown in 
Fig.~\ref{fig_modes} is of order $\approx 10^{-8}v_F/a$, to be compared with an analytical background of order 
$\approx 0.2 v_F/a$ (see the caption of Fig~\ref{fig_modes}).
These aspects make the numerical observation of the phenomenon extremely difficult, so that the effect of two 
very close EPs may be incorrectly interpreted as a mode-crossing.

\paragraph{Conclusions.} We have pointed out a  connection between properties of
graphene and the theory of non-Hermitian Hamiltonians, by showing that armchair graphene nanoribbons provide 
the first example of a nondissipative system described by a $\mathcal{PT}$-symmetric non-Hermitian Hamiltonian.
We have also established that the transverse momentum $p_y$ is an order parameter for the $\mathcal{PT}$ symmetry 
breaking. We have numerically verified the presence of exceptional points and shown that, in their neighborhood, 
the behavior of $p_x$ and of the eigenfunctions of the Dirac Hamiltonian, in the presence of a longitudinally 
invariant external potential, is theoretically well understood.


An aspect that certainly deserves further study is the effect of exceptional points on the transport properties of graphene nanoribbons, in the presence of a potential that varies also in the longitudinal direction. Moreover it would be interesting to study more in depth the properties of complex eigenmodes and the effects of the non-vanishing transverse momentum.

We thank M.~D'Elia for useful comments. 
D.~L., M.~M., and P.~M. gratefully acknowledge support from the EU FP7 IST Project GRAND (contract number 215752) via the IUNET consortium.

\end{document}